\documentclass[fleqn]{article}
\usepackage{amsmath,amssymb}
\usepackage{epsfig,graphicx}

\renewcommand{\theequation}{\arabic{section}.\arabic{equation}}
\renewcommand{\thesection}{\arabic{section}.}

\catcode`@=11 \@addtoreset{equation}{section} \catcode`@=12

\begin{document}
\title{\vskip-1.7cm \bf  Effective action and heat kernel in a toy model
of brane-induced gravity}
\date{}
\author{A.O.Barvinsky$^{1,\,2}$, A.Yu.Kamenshchik$^{3,\,4}$, C.Kiefer$^{5}$
and D.V.Nesterov$^{1}$} \maketitle \hspace{-8mm} {\,\,$^{1}$\em
Theory Department, Lebedev
Physics Institute, Leninsky Prospect 53, Moscow 119991, Russia\\
$^{2}$Sektion Physik, LMU, Theresienstr. 37, Munich, Germany\\
$^{3}$Dipartimento di Fisica and INFN, via Irnerio 46, 40126
Bologna, Italy\\
$^{4}$L.D.Landau Institute for Theoretical Physcis of Russian
Academy of Sciences, Kosygin str. 2, 119334 Moscow, Russia\\
$^{5}$Institut fur Theoretische Physik, Universit\"{a}t zu K\"{o}ln,
Z\"{u}lpicher Strasse 77, 50937 K\"{o}ln, Germany}
\begin{abstract}
We apply a recently suggested technique of the Neumann-Dirichlet
reduction to a toy model of brane-induced gravity for the
calculation of its quantum one-loop effective action. This model is
represented by a massive scalar field in the $(d+1)$-dimensional
flat bulk supplied with the $d$-dimensional kinetic term localized
on a flat brane and mimicking the brane Einstein term of the
Dvali-Gabadadze-Porrati (DGP) model. We obtain the inverse mass
expansion of the effective action and its ultraviolet divergences
which turn out to be non-vanishing for both even and odd spacetime
dimensionality $d$. For the massless case, which corresponds to a limit
of the toy DGP model, we obtain the Coleman-Weinberg type effective
potential of the system. We also obtain the proper time expansion of
the heat kernel in this model associated with the generalized
Neumann boundary conditions containing second order tangential
derivatives. We show that in addition to the usual integer and
half-integer powers of the proper time this expansion exhibits, depending
on the dimension $d$, either logarithmic terms or powers
multiple of one quarter. This property is considered in the context of
strong ellipticity of the boundary value problem, which can be
violated when the Euclidean action of the theory is not
positive definite.
\end{abstract}

\section{Introduction}
Modified theories of gravity in the form of braneworld models can in
principle account for the phenomenon of dark energy as well as for
nontrivial compactifications of multi-dimensional string models. It
becomes increasingly more obvious that one should include in such
models the analysis of quantum effects beyond the tree-level
approximation \cite{quantumDGP}. This is the only way to reach an
ultimate conclusion on the resolution of such problems as the
presence of ghosts \cite{ghosts} and low strong-coupling scale
\cite{scale}. Quantum effects in brane models are also important for
the stabilization of extra dimensions \cite{GarPujTan}, fixing the
cross-over scale in the Brans-Dicke modification of the DGP model
\cite{Pujolas} and in the recently suggested mechanism of the
cosmological acceleration generated by the four-dimensional
conformal anomaly \cite{slih}.

A general framework for treating quantum effective actions in brane models
(or, more generally, models with timelike and spacelike boundaries)
was recently suggested in \cite{BKRK,gospel,qeastb}. The main peculiarity
of these models is that due to quantum field fluctuations on the
branes the field propagator is subject to generalized Neumann
boundary conditions involving normal and tangential derivatives on
the brane/boundary surfaces. This presents both technical and
conceptual difficulties, because such boundary conditions are much
harder to handle than the simple Dirichlet ones. The method of
\cite{qeastb} provides a systematic reduction of
the generalized Neumann boundary conditions to
Dirichlet conditions. As a byproduct it disentangles from the quantum
effective action the contribution of the surface modes mediating the
brane-to-brane propagation (that is, within one brane),
which play a very important role in the
zero-mode localization mechanism of the Randall-Sundrum type
\cite{RS}. The main purpose of our paper here is
 to apply this method to a simplified toy
model of brane-induced gravity in order to demonstrate how it works
for the first nontrivial example of a field system with
boundary conditions involving second-order tangential derivatives.
As we will see, this model leads to qualitatively new structures
in the one-loop effective action, its renormalization, and the
associated heat kernel.

Briefly the method of \cite{qeastb} looks as follows. The action of
a (free field) brane model generally contains the bulk and the brane
parts,
    \begin{eqnarray}
    S[\,\phi\,]=\frac12\int_{\rm\bf B} d^{d+1}X\,\phi(X)\!
    \stackrel{\leftrightarrow}{F}\!(\nabla)\,\phi(X)
    +\frac12\int_{\rm \bf b}
    d^dx \,\varphi(x)\,
    \kappa(\partial)\,\varphi(x) \ ,                         \label{1}
    \end{eqnarray}
where the $(d+1)$-dimensional bulk and the $d$-dimensional brane
coordinates are labeled respectively by $X=X^A$ and $x=x^\mu$, and
the boundary values of bulk fields $\phi(X)$ on the brane/boundary
${\rm\bf b}=\partial\rm\bf B$ are denoted by $\varphi(x)$,
    \begin{eqnarray}
    \phi(X)\,\Big|_{\,\rm\bf b}=\varphi(x).        \label{2}
    \end{eqnarray}
The kernel of the bulk Lagrangian is given by the second order
differential operator $F(\nabla)$, whose derivatives
$\nabla\equiv\partial_X$ are integrated by parts in such a way that
they form bilinear combinations of first order derivatives acting on
two different fields (this is denoted in (\ref{1}) by the double-headed arrow).
Integration by parts in the bulk gives nontrivial surface terms on
the brane/boundary. In particular, this operation results in the
Wronskian relation for generic test functions $\phi_{1,\,2}(X)$,
    \begin{eqnarray}
    \int_{\rm\bf B} d^{\,d+1}X
    \left(\,\phi_1\stackrel{\rightarrow}{F}\!(\nabla)\phi_2-
    \phi_1\!\stackrel{\leftarrow}{F}\!(\nabla)\,\phi_2\right)=
    -\int_{\partial{\rm\bf B}} d^{\,d}x
    \left(\,\phi_1\stackrel{\rightarrow}{W}\!
    \phi_2-
    \phi_1\stackrel{\leftarrow}{W}\!
    \phi_2\right).                          \label{3a}
    \end{eqnarray}
Arrows everywhere here indicate the direction of action of
derivatives either on $\phi_1$ or $\phi_2$.

The brane part of the action contains as a kernel some local
operator $\kappa(\partial)$, $\partial=\partial_x$. Its order in
derivatives depends on the model in question. In the Randall-Sundrum
model \cite{RS}, for example, it is for certain gauges just an
ultralocal multiplication operator generated by the tension term on
the brane. In the Dvali-Gabadadze-Porrati (DGP) model \cite{DGP} this is a
second order operator induced by the brane Einstein term on the
brane, $\kappa(\partial)\sim\partial\partial/m$, where $m$ is the DGP
scale which is of the order of magnitude of the horizon scale,being responsible
for the cosmological acceleration \cite{Deffayet}. In the context of the
Born-Infeld action in D-brane string theory with vector gauge fields,
$\kappa(\partial)$ is a first-order operator \cite{open}.

In all these cases the variational procedure for the action
(\ref{1}) with dynamical (not fixed) fields on the boundary
$\varphi(x)$ naturally leads to generalized Neumann boundary
conditions of the form
    \begin{eqnarray}
    \left.\Big(\stackrel{\rightarrow}{W}\!(\nabla)
    +\kappa(\partial)\Big)\,\phi\,\right|_{\,\rm\bf b}
    =0,                                                     \label{3}
    \end{eqnarray}
which uniquely specify the propagator of quantum fields and,
therefore, a complete Feynman diagrammatic technique for the system
in question. The method of \cite{qeastb} allows one to
systematically reduce this diagrammatic technique to the one subject
to the Dirichlet boundary conditions $\phi|_{\,\rm\bf b}=0$. The
main additional ingredient of this reduction procedure is the brane
operator $\mbox{\boldmath$F$}^{\,\rm brane}(x,x')$ which is constructed
from the Dirichlet Green's function $G_D(X,X')$ of the operator
$F(\nabla)$ in the bulk,
    \begin{eqnarray}
    \mbox{\boldmath$F$}^{\,\rm brane}(x,x')=-
    \stackrel{\rightarrow}{W}\!(\nabla_X\!)\,G_{D}(X,X')\!
    \stackrel{\leftarrow}{W}\!(\nabla_{X'}\!)
    \,\Big|_{\,X=e(x),\,X'=e(x')}
    +\kappa(\partial)\,\delta(x,x')\ .           \label{5}
    \end{eqnarray}
This expression expresses the fact that the kernel of the Dirichlet Green's
function is being acted upon both arguments by the Wronskian
operators with a subsequent restriction to the brane, with $X=e(x)$
denoting the brane embedding function.

As shown in \cite{qeastb}, this operator determines the
brane-to-brane propagation of the physical modes in the system with
the classical action (\ref{1}) (its inverse is the brane-to-brane
propagator) and additively contributes to its full one-loop
effective action according to
    \begin{eqnarray}
    \mbox{\boldmath$\varGamma$}_{\rm 1-loop}\equiv\frac12\;
    {\rm Tr}_N^{(d+1)}\ln F=\frac12\;{\rm Tr}_D^{(d+1)}\ln F
    +\frac12\;{\rm Tr}^{(d)}
    \ln \mbox{\boldmath$F$}^{\,\rm brane},    \label{11}
    \end{eqnarray}
where ${\rm Tr}_{D,\,N}^{(d+1)}$ denotes functional traces of the
bulk theory subject to Dirichlet and Neumann boundary conditions,
respectively, while ${\rm Tr}^{(d)}$ is a functional trace in the
boundary $d$-dimensional theory. The full quantum effective action
of this model is obviously given by the functional determinant of
the operator $F(\nabla)$ subject to the generalized Neumann boundary
conditions (\ref{5}), and the above equation reduces its calculation
to that of the Dirichlet boundary conditions plus the contribution
of the brane-to-brane propagation.

Here we apply (\ref{11}) to a simple model of a scalar field which
mimicks in particular the properties of the brane-induced
gravity models and the DGP model \cite{DGP}. This
is the $(d+1)$-dimensional massive scalar field $\phi(X)=\phi(x,y)$
with mass $M$ living in the half-space $y\geq 0$ with the additional
$d$-dimensional kinetic term for $\varphi(x)\equiv\phi(x,0)$
localized at the brane (boundary) at $y=0$,
    \begin{eqnarray}
    S[\,\phi\,]=\frac12\int\limits_{y\geq 0}
    d^{d+1}X\,\Big((\nabla\phi(X))^2
    +M^2\phi^2(X)\Big)
    +\frac1{2m}\int
    d^dx \,(\partial\varphi(x))^2.           \label{1.1}
    \end{eqnarray}
Here and in what follows we work in a flat Euclidean
(positive-signature) spacetime. Therefore, this action corresponds
to the following choice of $F(\nabla)$ in terms of
$(d+1)$-dimensional and $d$-dimensional D'Alembertians (Laplacians)
    \begin{eqnarray}
    &&F(\nabla)=M^2-\Box^{(\,d+1)}=
    M^2-\Box-\partial_y^2,\,\,\,
    \Box=\Box^{(d)}\equiv\partial_\mu^2.              \label{6}
    \end{eqnarray}
Its Wronskian operator is given by the normal derivative with
respect to outward-pointing normals to the brane, $W=-\partial_y$,
and the boundary operator $\kappa(\partial)$ equals
    \begin{eqnarray}
    \kappa(\partial)=-\frac1m\,\Box \ ,            \label{8}
    \end{eqnarray}
where the dimensional parameter $m$ mimicks the role of the DGP
scale \cite{DGP}. Thus, the generalized Neumann boundary conditions
in this model involve second-order derivatives tangential to the
brane,
    \begin{eqnarray}
    \Big(\partial_y
    +\frac1m\,\Box\Big)\,\phi\,
    \Big|_{\,\rm\bf b}=0,               \label{1.3}
    \end{eqnarray}
cf. (\ref{3}) with $W=-\partial_y$ and $\kappa=-\Box/m$.

As we show below, the brane-to-brane operator for such a model has
the form of the following pseudodifferential operator,
    \begin{eqnarray}
    \mbox{\boldmath$F$}^{\,\rm brane}(\partial)
    =\frac1m\,(-\Box+m\sqrt{M^2-\Box}).          \label{9}
    \end{eqnarray}
In the massless case of the DGP model \cite{DGP}, $M=0$, this
operator is known to mediate the gravitational interaction on
the brane, interpolating between the four-dimensional Newtonian law at
intermediate distances and the five-dimensional law at the horizon
scale $\sim 1/m$ \cite{scale}. We calculate the effective action
(\ref{11}) for this model both in the form of the $1/M$-expansion
and exactly in terms of a special hypergeometric function
representation (as a function of $M$ and the dimensionless ratio
$m/M$) and find its ultraviolet divergences.

As a byproduct of the Neumann to Dirichlet reduction (\ref{11}), the
technique of \cite{qeastb} also yields a method for obtaining the
proper time expansion for the heat kernel associated with the
boundary conditions (\ref{3}). For simple Neumann (Robin) boundary
conditions containing at most first order derivatives tangential to
the boundary, this expansion has the form
    \begin{eqnarray}
    {\rm Tr}^{(\,d+1)}\,e^{-sF(\nabla)}=
    \frac1{(4\pi s)^{(d+1)/2}}\,
    \sum\limits_{n=0}^\infty
    \left(\,s^n\,A_n+s^{n/2}\,B_{n/2}\,\right).   \label{10}
    \end{eqnarray}
In addition to the well-known bulk integrals $A_n$ of the local
Schwinger-DeWitt coefficients of integer powers of the proper time
\cite{DeWitt,DeWitt1,PhysRep}, it contains surface integrals
$B_{n/2}$ as coefficients of both integer and half-integer powers of
$s$
\cite{McKean-Singer,BransGilkey,Osborn-McAvity,BransGilkeyVas,Vassilevich}.
They are sufficiently easy to calculate for the Dirichlet boundary
conditions \cite{McKean-Singer,Vassilevich}, but become much harder
to obtain for the Robin and generalized Neumann case with a
growing number of tangential derivatives
\cite{Osborn-McAvity,DowkerKirsten,AvramEsp}. As shown in
\cite{qeastb}, the Neumann-Dirichlet reduction method
simplifies their calculation essentially. For second-order derivatives they are
not known at all, and a toy DGP model of the above type seems to be
the first application of the heat kernel mehtod subject to the boundary
conditions (\ref{1.3}).

It turns out that in the case of (\ref{1.3}) even the very structure
(\ref{10}) is incorrect, because for even $d$ it contains also terms
logarithmic in $s$ and for odd $d$ it has also powers of $s$ which
multiples of a quarter. We calculate these additional terms in the
heat kernel expansion, discuss their relation to a nontrivial
analytic structure of the brane part of the effective action
(\ref{11}) and also to the problem of strong ellipticity
\cite{ellipticity} of the boundary value problem (\ref{1.3}). In
particular, the latter is shown to be determined by the positivity
of the action (\ref{1.1}) or the positive-definiteness of the brane
operator (\ref{9}).

The paper is organized as follows. In Sect.~2 we derive the
Dirichlet and brane parts of the effective action (\ref{11}).
Sects.~3 and 4 present its inverse mass expansion and ultraviolet
divergences in various dimensions. In Sect.~5 we obtain a new type
of the proper time expansion for the heat kernel associated with the
boundary conditions involving second-order tangential derivatives.
In Sect.~6 we present the hypergeometric function representation of
the effective action and analyze the limit of a simple Neumann
boundary condition corresponding to $m\to\infty$; we also consider
the massless limit $M=0$ which gives the effective potential in the
toy DGP model. In the Conclusions we discuss these results in the
context of a possible violation of strong ellipticity for the
boundary conditions (\ref{1.3}), their potential applications in
braneworld models including gravitation and the use of the proper
time method in brane models. Three appendices contain the derivation
of the inverse mass expansion of the effective action, its exact
hypergeometric function representations, and the presentation of the
status of the strong ellipticity problem in a toy DGP model.

\section{Dirichlet and brane-to-brane contributions}We begin by
constructing the Dirichlet part of the effective action. The basic
Dirichlet Green's function of the model can be obtained by the
proper time integration of the corresponding heat equation kernel,
    \begin{eqnarray}
    &&K_D(s\,|X,X')=e^{s\Box_{(d+1)}}\delta(X,X').     \label{2.1}
    \end{eqnarray}
It follows by the method of images from its well-known expression in
the flat $(d+1)$-dimensional spacetime without boundaries
    \begin{eqnarray}
    &&K_D(s\,|X,X')=
    \frac1{(4\pi s)^{\frac{d+1}2}}\,
    \left\{\,\exp\left(-\frac{(x-x')^2+(y-y')^2}
    {4s}\right)\right.\nonumber\\
    &&\qquad\qquad\qquad\qquad
    \qquad\qquad\qquad\qquad
    \left.-\exp\left(-\frac{(x-x')^2+(y+y')^2}
    {4s}\right)\,\right\}.                        \label{HKD}
    \end{eqnarray}

The functional trace of this heat kernel contains two terms --- bulk
and boundary integrals of the only two nonvanishing Schwinger-DeWitt
coefficients,
    \begin{eqnarray}
    &&{\rm Tr}^{(\,d+1)}_D\,e^{s\,\Box_{(d+1)}}=
    \int_{y\geq 0} d^{d+1}X\,K_D(s|X,X)\nonumber\\
    &&\qquad\qquad\qquad\qquad\qquad
    =\frac1{(4\pi s)^{\frac{d+1}2}}\,
    \left(\,\int_{y\geq 0} d^{d+1}X
    -\frac{\sqrt\pi}2\,s^{1/2}\int d^dx\right).   \label{TrKD}
    \end{eqnarray}
The corresponding Dirichlet-type effective action for the model with
the mass $M$ in the bulk can be obtained by the following proper
time integration,
    \begin{eqnarray}
    &&\frac12\,{\rm Tr}_{D}\ln\,\Big[-\Box_{(d+1)}+M^2\,\Big]
    =-\frac12\,{\rm Tr}_{D}\int_0^\infty
    \frac{ds}s\;e^{s\,\Box_{(d+1)}-s\,M^2}\nonumber\\
    &&\nonumber\\
    &&\qquad\qquad
    =-\frac12\Big(\,\frac{M^2}{4\pi}\Big)^{\!\frac{d+1}2}\,
    \Gamma\Big(\!-\frac{d+1}2\Big)\!\int\limits_{y\geq 0} d^{d+1}X+
    \frac18\Big(\,\frac{M^2}{4\pi}\Big)^{\!\frac{d}2}
    \Gamma\Big(\!-\frac{d}2\Big)
    \,\int d^dx.              \label{2.5}
    \end{eqnarray}
The dimensionality $d$ will be treated as a parameter of the
dimensional regularization. Therefore, this expression contains
ultraviolet divergences as the poles of Gamma functions at negative
integer values of their arguments.  These divergences are
represented here either by the bulk or boundary surface integrals,
depending on whether the total spacetime dimensionality $(d+1)$ is
even or odd.

For the brane part of the action (\ref{11}) we need the
brane-to-brane operator (\ref{5}) which is based on the Dirichlet
Green's function of the model. The latter is also exactly calculable
in elementary functions because the corresponding proper time
integral can be expressed in terms of the modified Bessel function of
half-integer order,
    \begin{eqnarray}
    &&G_{D}(X,X')=\int\limits_0^\infty ds\,
    K_D(s\,|X,X')\,e^{-sM^2}\nonumber\\
    &&\qquad\qquad\qquad
    =\int\limits_0^\infty \frac{ds}{\sqrt{4\pi s}}\,
    \left(e^{-(y-y')^2/4s}-e^{-(y+y')^2/4s}\right)\,
    e^{s(\Box-M^2)}\delta(x,x')\nonumber\\
    &&\qquad\qquad\qquad=\frac1{2\sqrt{M^2-\Box}}\,
    \left(e^{-|\,y-y'|\,\sqrt{M^2-\Box}}
    -e^{-(y+y')\,\sqrt{M^2-\Box}}\,\right)\,\delta(x,x').  \label{2.6a}
    \end{eqnarray}
Therefore the first term of (\ref{5}) takes the form of a
square-root operator \cite{qeastb},
    \begin{eqnarray}
    \stackrel{\rightarrow}{\partial_y}G_{D}(X,X')
    \stackrel{\leftarrow}{\partial_y}\!
    \,\Big|_{\,X=(x,0),\,Y=(x',0)}=\sqrt{M^2-\Box}\,\delta(x,x')\ ,
    \end{eqnarray}
and the full brane-to-brane operator (\ref{5}) is given by
(\ref{9}).

The operator (\ref{9}) is of a nonlocal pseudodifferential nature,
and no conventional proper time representation is known for its
functional determinant (see Sect.7, though). Therefore we will
calculate the latter in the basis of Fourier modes --- the
eigenmodes of $\mbox{\boldmath$F$}^{\,\rm brane}$. By resolving the
$d$-dimensional delta-function in the Fourier integral we have
    \begin{eqnarray}
    &&\frac12\;{\rm Tr}^{(d)}
    \ln \mbox{\boldmath$F$}^{\,\rm brane}=\frac12\,\int d^dx\,
    \ln\left(-\Box+m \sqrt{M^2-\Box}\,\right)\,
    \delta(x,x')\,\Big|_{x'=x}\nonumber\\
    &&\qquad\qquad\qquad\quad
    =\frac12\,\int d^dx\,\frac1{(2\pi)^d}\int d^dp\,\ln
    \left(p^2+m\sqrt{M^2+p^2}\,\right)\nonumber\\
    &&\qquad\qquad\qquad\quad
    =\frac1{(4\pi)^{d/2} \varGamma(d/2)}\int d^d x\,
    \int_0^\infty dp\,p^{d-1}
    \,\ln\left(p^2+m\sqrt{M^2+p^2}\,\right),  \label{2.6}
    \end{eqnarray}
where $p$ is the radial integration variable in the momentum space,
    \begin{eqnarray}
    p=\sqrt{p_\mu p^\mu}.
    \end{eqnarray}

As we see, the mass parameter $M^2$ enters the
logarithmic function here in a very nontrivial way,
 so a typical $1/M^2$-expansion of the
local Schwinger-DeWitt expansion is far from being straightforward.
In the next section we derive this expansion by converting the expression
(\ref{2.6}) into the form of a so-called integral with a weak
singularity to which a known asymptotic expansion technique can be
directly applied.

\section{Inverse mass expansion}
By integrating in (\ref{2.6}) by parts and using the rules of the
dimensional regularization, which annihilates purely power-divergent
integrals, we have
    \begin{eqnarray}
    \int\limits_0^\infty dp\,p^{d-1}
    \,\ln
    \left(p^2+m\sqrt{M^2+p^2}\,\right)
    =-\frac1d\int_0^\infty dp\,
    p^{d+1}\,\frac{2+m/\sqrt{M^2+p^2}}
    {p^2+m\sqrt{M^2+p^2}}.                \label{3.1}
    \end{eqnarray}
Then, with the change of the integration variable from $p$ to $t$,
    \begin{eqnarray}
    t=\frac{p^2}{2M\sqrt{M^2+p^2}},\,\,\,
    p=M\left[2t\,(\sqrt{1+t^2}+t)\right]^{1/2},  \label{3.2}
    \end{eqnarray}
the integral takes the form
    \begin{eqnarray}
    &&\int\limits_0^\infty dp\,p^{d-1}
    \,\ln \left(p^2+m\sqrt{M^2+p^2}\,\right)=
    \frac{(2M^2)^{d/2}}d\,\varepsilon\,I \ ,   \label{3.3}
    \end{eqnarray}
where
    \begin{eqnarray}
    &&I=\int_0^\infty dt\,
    t^{d/2-1}\,(t+\varepsilon)^{-1}\,
    \varphi(t),                       \label{3.4}\\
    &&\varphi(t)=
    (\sqrt{1+t^2}+t)^{d/2},            \label{3.4a}
    \end{eqnarray}
and
    \begin{eqnarray}
    &&\varepsilon=\frac{m}{2M}.              \label{3.6}
    \end{eqnarray}

Obviously the $1/M$ asymptotic expansion corresponds to the
asymptotic expansion in $\varepsilon\to 0$ --- the limit in which
the power-law singularity of the integrand occurs at the lower
integration limit $t=0$. Here the integrand is not analytic because
of the factor $t^{d/2-1}$ having a branch point at $t=0$. Remember
that the integral should be calculated for a generic dimensionality
$d$ that should be analytically continued to the complex plane in
order to regularize the ultraviolet divergences appearing at the
upper integration limit $t\to\infty$. Therefore, one should expect
that the expansion of the integral will also have a part nonanalytic
in $\varepsilon\to 0$.

As shown in Appendix A by the asymptotic expansion method for
integrals with a weak singularity \cite{Fedoryuk}, this expansion
indeed has the form
    \begin{eqnarray}
    I=\frac{\pi\,\varepsilon^{d/2-1}}{\sin(\pi d/2)}\,
    \varphi(-\varepsilon)+
    \sum_{j=0}^\infty a_{2j}\,
    \varepsilon^{2j},        \label{3.14one}
    \end{eqnarray}
where the first term --- the nonanalytic part --- has a branch-point
singularity. Interestingly, the coefficient of this nonanalytic
factor is exactly expressed through the function (\ref{3.4a})
itself but with the flipped sign of the argument. Its expansion in
higher powers of $\varepsilon$, given by Eq.~(\ref{3.14}) in Appendix
A, is determined by the derivatives $\varphi^{(n)}(0)\equiv
d^n\varphi/dt^n(0)$ which explicitly equal
    \begin{eqnarray}
    &&\varphi^{(n)}(0) = 2^{n-2}(-1)^{n+1}\frac{d\,
    \Gamma\left(\frac{n}{2}
    -\frac{d}{4}\right)}{\Gamma\left(1-\frac{n}{2}
    -\frac{d}{4}\right)}\nonumber \\
    &&\qquad\qquad\quad=\frac{d}{2}\left(\frac{d}{2}
    -(n-2)\right)\,\left(\frac{d}{2}
    -(n-4)\right)\times\cdots\times
    \left(\frac{d}{2}-(2-n)\right).            \label{3.15}
    \end{eqnarray}
The analytic part of (\ref{3.14one}) contains only even powers of
$\varepsilon$, with the coefficients
    \begin{eqnarray}
    a_{2j}=-2^{2j-1-d/2}\,d\,
    \frac{\Gamma(j-\frac{d-1}2)\,
    \Gamma(\frac{d}2-1-2j)}{\Gamma(-j+\frac12)}.     \label{3.18}
    \end{eqnarray}

Thus, finally, on account of (\ref{2.6}) and (\ref{3.3}) the inverse
mass expansion for the brane part of the action takes the form of
two terms having qualitatively different analytic behaviour in the mass
parameters $M$ and $m$,
    \begin{eqnarray}
    &&\frac12\;{\rm Tr}^{(d)}
    \ln \mbox{\boldmath$F$}^{\,\rm brane}\nonumber\\
    &&\quad\quad=
    \!
    \frac1{d\,\Gamma(d/2)}\int
    d^dx \left[\frac\pi{\sin{\frac{\pi d}2}}
    \left(\frac{Mm}{4\pi}\right)^{\!d/2}\!\!
    \varphi(-\varepsilon)+\left(\frac{M^2}{2\pi}\right)^{\!d/2}
    \sum_{j=0}^\infty a_{2j}\,
    \varepsilon^{2j+1}
    \right].                       \label{3.20}
    \end{eqnarray}

\section{Ultraviolet divergences versus spurious infrared poles}
The integral (\ref{3.4}) has a power divergence of order $d-1$ at the
upper integration limit. Its differentiation with respect to
$\varepsilon$ improves the convergence of the integral which becomes
ultraviolet finite after $d$ differentiations. This means that the
ultraviolet divergence of (\ref{3.4}) is a polynomial in
$\varepsilon$ of order $d-1$. The
ultraviolet divergence of the brane action
is a polynomial of order $d$ in $m$ and
$M$, respectively, cf. Eq.~(\ref{3.3}). In dimensional regularization these
divergences manifest themselves as poles in the dimensionality $d$
analytically continued to the complex plane. From (\ref{3.20}) it
follows, however, that for even $d$ the nonanalytic and analytic
parts of the inverse mass expansion separately have poles to all
orders in $\varepsilon$ --- the poles of $\pi/\sin(\pi d/2)$ and the
poles of one of the Gamma functions in the numerator of (\ref{3.18})
for all $2j\geq d/2-1$. This implies an intrinsic cancelation between
the infinite sequence of poles in the nonanalytic and analytic parts
of (\ref{3.20}). This cancelation of spurious poles, which have the
nature of infrared divergences, can be directly observed by
calculating separately these two contributions.

The real ultraviolet divergences of (\ref{3.4}) can be independently
obtained by means of integration by parts. In terms of the new
integration variable $x=1/t$ the integral becomes divergent at the
lower integration limit,
    \begin{eqnarray}
    &&I = \int_{0}^{\infty} dx x^{-d}f(x),         \label{ultrav1}\\
    &&f(x)\equiv (1+\varepsilon x)^{-1}
    (\sqrt{1+x^2}+1)^{\frac{d}{2}}.               \label{f-define}
    \end{eqnarray}
We call in the following the physical dimension of spacetime $N$ in
order to distinguish it from the formal dimension used in the
integrals.
Analytically continuing the spacetime dimensionality from its
physical value $N$ to the domain of convergence, where integrations
by parts are possible without introducing extra surface terms, we
have
    \begin{eqnarray}
    I = \frac{1}{(d-1)(d-2)
    \cdots(d-N)}\int_0^{\infty} dx
    x^{N-d} f^{(N)}(x).                                \label{byparts}
    \end{eqnarray}
Then taking the limit to the real physical dimensionality,
    \begin{eqnarray}
    d=N-\delta,\,\,\,\,\delta\to +0,
    \end{eqnarray}
we have the logarithmic divergence of this integral as a residue of
the pole in $\delta\to 0$,
    \begin{eqnarray}
    I^{\rm div} =
    \frac{1}{\delta}\frac{f^{(N-1)}(0)}{\Gamma(N)}.    \label{ultrav-der}
    \end{eqnarray}

By applying this formula we obtain the ultraviolet divergences which for
even and odd dimensionalities look as follows. For even $d$ being a multiple
of 4, $d = 4k-\delta,\  k=0,1,\cdots$, the divergences read
    \begin{eqnarray}
    &&\frac12\;{\rm Tr}^{(d)}
    \ln \mbox{\boldmath$F$}^{\,\rm brane}
    \Big|^{\rm div}\nonumber\\
    &&\quad=
    -\frac{1}{\delta}\!\times\frac1{2(4\pi)^{d/2}
    \Gamma\left(\frac{d}{2}\right)}
    \int d^{4k}x\sum_{j=k-1}^{2k-1}
    \frac{\Gamma(j+1)\,M^{4k-2j-2}\,m^{2j+2}}
    {\Gamma(2k-j)(2j+2-2k)!}.      \label{ultra-der-even0}
    \end{eqnarray}
For even $d$ being a multiple of 2, $d=4k+2-\delta,\ k=0,1,\cdots$, they
have the form
    \begin{eqnarray}
    &&\frac12\;{\rm Tr}^{(d)}
    \ln \mbox{\boldmath$F$}^{\,\rm brane}
    \Big|^{\rm div}\nonumber\\
    &&\quad=
    -\frac{1}{\delta}\!\times\frac1{2(4\pi)^{d/2}
    \Gamma\left(\frac{d}{2}\right)}
    \int d^{4k+2}x\sum_{j=k}^{2k}
    \frac{\Gamma(j+1)\,M^{4k-2j}\, m^{2j+2}}
    {\Gamma(2k+1-j)(2j+1-2k)!}.          \label{ultra-der-even2}
    \end{eqnarray}
Similarly, for odd $d=2k+1-\delta$ the ultraviolet divergences of
the brane effective action read
    \begin{eqnarray}
    &&\frac12\;{\rm Tr}^{(d)}
    \ln \mbox{\boldmath$F$}^{\,\rm brane}\Big|^{\rm div}\nonumber\\
    &&\quad=
    \frac{1}{\delta}\!\times\frac1{2(4\pi)^{d/2}
    \Gamma\left(\frac{d}{2}\right)}\int d^{2k+1}x\,\sum_{j=0}^{k}
    \frac{(-1)^{k+j+1}\Gamma\left(k-2j-\frac12\right)}
    {\Gamma\left(\frac{1}{2}-j\right)\,(k-j)!}\,
    M^{2k-2j}\, m^{2j+1}.                     \label{diver-odd}
    \end{eqnarray}

These results can be directly checked by disentangling the poles in
spacetime dimensionality in the general inverse mass expansion
(\ref{3.20}). This calculation confirms the cancelation of the
fictitious infrared divergences in even $d$ mentioned above.
Although these divergences have a spurious nature and only arise at
intermediate calculational stages, their presence reflects a
nontrivial analytic structure of the asymptotic mass expansion
(\ref{3.20}). As we will see below, they entail a nontrivial form of
the heat kernel expansion corresponding to the generalized Neumann
boundary conditions with second-order derivatives tangential to the
boundary.

\section{Heat kernel expansion}
Because of the nontrivial pseudodifferential nature of the
brane-to-brane operator (\ref{9}) and the inverse mass expansion of
its action (\ref{3.20}) the heat kernel for the generalized Neumann
boundary conditions (\ref{1.3}) does not have a typical expansion in
integer and half-integer powers of the proper time (\ref{10}). Thus
we assume a more general structure of this expansion in the form
    \begin{eqnarray}
    &&{\rm Tr}^{(\,d+1)}_N\,e^{s\Box_{(d+1)}}=
    \frac1{(4\pi s)^{\frac{d+1}2}}\,
    \left(\,\int_{y\geq 0} d^{d+1}X+\sum\limits_{\{p\}}
    s^p\,\int d^dx\,b_{\{p\}}\,\right),                    \label{HK1}
    \end{eqnarray}
where the summation runs over some unknown set of powers $\{p\}$ of
the proper time with some unknown coefficients of the surface
integrals $b_{\{p\}}$. The bulk integral here involves only one term
corresponding to the only non-vanishing bulk Schwinger-DeWitt
coefficient $a_0(X,X)=1$ (which is independent of the boundary
conditions and coincides with the one in spacetime without
boundaries).

This expansion generates the inverse mass expansion for the
effective action of the problem (\ref{1}),
    \begin{eqnarray}
    &&\frac12\,{\rm Tr}_{N}\ln\,\Big[\,M^2-\Box_{(d+1)}\Big]
    =-\frac12\,{\rm Tr}_{N}\int_0^\infty
    \frac{ds}s\;e^{s\,\Box_{(d+1)}-s\,M^2}\nonumber\\
    &&\nonumber\\
    &&\quad
    =-\frac12\Big(\,\frac{M^2}{4\pi}\Big)^{\!\frac{d+1}2}\,\left\{\,
    \Gamma\Big(\!-\frac{d+1}2\Big)\!\int\limits_{y\geq 0} d^{d+1}X+
    \sum_{\{p\}}
    \frac{\Gamma\left(p-\frac{d+1}2\right)}{M^{2p}}\!
    \int d^dx\,b_{\{p\}}\right\}.                         \label{5.2}
    \end{eqnarray}
Our goal now will be to determine the range of summation  $\{p\}$
and the coefficients $b_{\{p\}}$ by comparing this expression with
the inverse mass expansion of (\ref{11}) known from (\ref{2.5}) and
(\ref{3.20}).

We begin by considering the most interesting case of even
dimensionality which under dimensional regularization reads as
$d=2k-\delta$, $\delta\to 0$. Even though the heat kernel is an
ultraviolet finite object (which generates UV divergences in the
action due to the divergence of the proper time integration), we
need this regularization to regulate intermediate infrared
divergences which cancel out in the final answer. Thus, assembling
together (\ref{2.5}) and (\ref{3.20}) we get
    \begin{eqnarray}
    &&\frac12\,{\rm Tr}_{N}\ln\,\Big[\,M^2-\Box_{(d+1)}\Big]
    =-\frac12\,\Big(\frac{M^2}{4\pi}\,\Big)^{\!\frac{d+1}2}\,
    \Gamma\Big(\!-\frac{d+1}2\Big)\,
    \int_{y\geq 0} d^{d+1}X\nonumber\\
    &&\nonumber\\
    &&\qquad\qquad
    +\left(\frac{M^2}{4\pi}\right)^{\!\frac{d}2}\int
    d^dx\,\left(\;\frac18\,\Gamma(-d/2)+
    \sum_{j=0}^\infty
    \,\frac{2^{\frac{d}2-2j-1}}{d\,\Gamma(\frac{d}2)}\;a_{2j}\,
    \frac{ m^{2j+1}}{M^{2j+1}}\right.\nonumber\\
    &&\nonumber\\
    &&\qquad\qquad
    \left.-\frac12\,\Gamma(\!-d/2)\sum_{n=k}^\infty\,
    \Big(\!-\frac12\,\Big)^{n-k}\,
    \frac{\varphi^{(n-k)}(0)}{(n-k)!}\,
    \frac{ m^{n-\delta/2}}{M^{n-\delta/2}}\,\right),\,\,\,\,
    d=2k-\delta.                                             \label{5.3}
    \end{eqnarray}
Comparing it with (\ref{5.2}) we immediately get the range of
summation over the proper-time powers $\{p\}$ which we will label by the
integer numbers $j$ and $n$,
    \begin{eqnarray}
    \{p\}=\left\{\begin{array}{ll}\,1/2,\,\,\,&\\
    \,j,\,\,\,&\,j\geq 1, \\
    \,n/2-\delta/4,\,\,\,&\,n\geq k+1,\end{array}\right.
    \end{eqnarray}
and the corresponding coefficients
    \begin{eqnarray}
    b_{\{p\}}=\left\{\begin{array}{ll}\,b_{1/2},\,\,\,&\\
    \,b_{j},\,\,\,&\,j\geq 1, \\
    \,b_{n/2-\delta/4}\equiv{\tilde b}_{n/2} ,
    \,\,\,&\,n\geq k+1.\end{array}\right.
    \end{eqnarray}

In terms of $b_{1/2}$, $b_j$ and $\tilde b_{n/2}$ the heat kernel
trace takes the form
    \begin{eqnarray}
    &&{\rm Tr}^{(\,d+1)}_N\,e^{s\,\Box_{(d+1)}}=
    \frac1{(4\pi s)^{\frac{d+1}2}}\,
    \int\limits_{y\geq 0} d^{d+1}X\nonumber\\
    &&\qquad\qquad\quad
    +\frac1{(4\pi s)^{\frac{d+1}2}}\,
    \int d^{d}x
    \left\{\;b_{1/2}\,s^{1/2}+\sum_{j=1}^\infty b_{j}\,s^{j}+
    \sum_{n=k+1}^\infty
    \tilde b_{n/2}\,s^{(n-k)/2+d/4}\right\}.             \label{HK2}
    \end{eqnarray}
The lowest order surface coefficient coincides with the Dirichlet
one,
    \begin{eqnarray}
    b_{1/2}=b_{1/2}^D\equiv-\frac{\sqrt\pi}2,              \label{b1/2}
    \end{eqnarray}
whereas the higher order ones take the following form in view of the
known expressions (\ref{3.15}) and (\ref{3.18}) for the coefficients
of the nonanalytic and analytic parts of the action,
    \begin{eqnarray}
    &&b_{j}=\sqrt\pi\,
    \frac{\Gamma(k-2j+1-\frac{\delta}2)}
    {\Gamma(\frac{d}2)\,
    \Gamma(\frac{3}2-j)}\,m^{2j-1},\,\,\,j\geq 1,        \label{bodd}\\
    &&{\tilde b}_{n/2}=\sqrt\pi\,
    \frac{\Gamma(1-k+\frac{\delta}2)}
    {\Gamma(\frac{3-n}2+\frac\delta4)}\,
    \frac{1}{(n-1-k)!}
    \,m^{n-1-\delta/2},\,\,\,n\geq k+1.
    \end{eqnarray}

The first few coefficients $b_j$ are ultraviolet finite,
    \begin{eqnarray}
    b_{j}\to \sqrt\pi\,\frac{\Gamma(k-2j+1)}
    {\Gamma(k)\,\Gamma(\frac{3}2-j)}\,m^{2j-1},\,\,\,
    \delta\to 0,\,\,\,2\leq 2j\leq k.
    \end{eqnarray}
However, for $2j>k$ they are divergent,
    \begin{eqnarray}
    b_j=-2^{4-2j}(-1)^{j+k}
    \frac{\Gamma(2j-2)}{\Gamma(j-1)\,
    \Gamma(2j-k)\,\Gamma(k)}\,m^{2j-1}\,\frac1\delta+...,
    \,\,\,\delta\to 0,
    \end{eqnarray}
but the complimentary coefficients $\tilde b_j$ have poles with residues
that are exactly opposite in sign, $(b_j+\tilde b_j)^{\rm pole}=0$.
This is certainly a manifestation of the cancellation of infrared
divergences between the analytic part of (\ref{3.20}) related to
$b_j$ and the non-analytic part related to $\tilde b_j$. As a result
the heat kernel stays well defined but acquires logarithmic terms in
$s$ because
    \begin{eqnarray}
    &&b_{j}\,s^{j}+
    {\tilde b}_{j}\,s^{j-\delta/4}=
    s^j\,\Big(\,\beta_j\,\ln(sm^2)+\gamma_j\,\Big),       \\
    &&\beta_j=-2^{2-2j}(-1)^{j+k}
    \frac{\Gamma(2j-2)}{\Gamma(j-1)\,\Gamma(2j-k)\,
    \Gamma(k)}\,m^{2j-1}.
    \end{eqnarray}
The rest of the coefficients with half-integer numbers $\tilde
b_{j+1/2}$ are finite,
    \begin{eqnarray}
    \tilde b_{j+1/2}=\sqrt\pi\,(-1)^{k+j}
    \frac{\Gamma(j)}{2\Gamma(2j+1-k)\,\Gamma(k)}\,m^{2j}.
    \end{eqnarray}

Thus finally the heat kernel trace in even integer (unregulated)
dimensionality of a brane $d=2k$ takes the form
    \begin{eqnarray}
    &&{\rm Tr}^{(\,d+1)}_N\,e^{s\,\Box_{(d+1)}}=
    \frac1{(4\pi s)^{\frac{d+1}2}}\,
    \int\limits_{y\geq 0} d^{d+1}X\nonumber\\
    &&\qquad\qquad\qquad\quad
    +\frac1{(4\pi s)^{\frac{d+1}2}}\,
    \int d^{d}x \left\{\;b_{1/2}\,s^{1/2}
    +\sum_{j=1}^{[\,k/2\,]} b_{j}\,s^{j}+
    \sum_{2j\geq k}^\infty \tilde b_{j+1/2}\,s^{j+1/2}
    \right.\nonumber\\
    &&\qquad\qquad\qquad\qquad\qquad\qquad\qquad\qquad
    +\left.
    \sum_{2j\geq k+1}^\infty
    \Big(\beta_j\,\ln(sm^2)+\gamma_j\Big)\,s^j
    \,\right\}.                                      \label{HK3}
    \end{eqnarray}

In odd dimensions, $d=2k+1$, the poles of infrared nature are absent,
so that both $b_j$ and $\tilde b_{n/2}$ are finite, and no
logarithmic terms arise in the heat kernel. One can check that its
trace has explicitly the form (\ref{HK2}) with $d=2k+1$. Thus it has
not only half-integer powers of the proper time, but also powers
multiple of a quarter.

Note that $b_{1/2}$ for all {\em finite} $m$ has the value
(\ref{b1/2}) characteristic of the Dirichlet problem. At the same
time, for $m\to\infty$ our problem reduces to the case of Neumann
conditions, cf. Eq.(\ref{1.3}), corresponding to
$b_{1/2}^N=+\sqrt\pi/2$ \cite{McKean-Singer}. However, this does not
present any contradiction, because the asymptotic expansion in $1/M$
is obviously not homogeneous in $m\to\infty$, as it involves growing
positive powers of $m$. To analyze the limit of large $m$ or
$\varepsilon\to\infty$, which includes the case of the massless DGP
model, $M=0$, we have to consider another representation of the
effective action. This is discussed in the next section.

\section{Massless limit:  effective potential in a toy DGP model}
As shown in Appendix B, the brane part of the effective action
(\ref{2.6}) can be exactly represented in terms of the
hypergeometric function. This representation has the form
    \begin{eqnarray}
    &&\frac12\;{\rm Tr}^{(d)}
    \ln \mbox{\boldmath$F$}^{\,\rm brane}=
    \frac1{d\,\Gamma(d/2)}\int
    d^dx \,\frac\pi{\sin{\frac{\pi d}2}}
    \left(\frac{Mm}{4\pi}\right)^{\!d/2}\!\!
    \varphi(-\varepsilon)\nonumber\\
    &&\qquad\qquad\qquad
    -\left(\frac{M^2}{4\pi}\right)^{\!d/2}\,
    \frac{
    \Gamma\big(\,\frac{1-d}2\,\big)}{2d\,\sqrt\pi}
    \,\int
    d^dx \sum\limits_{v\,=\,v_\pm}\,
    v\,F\Big(1,\,\frac{1-d}2;\,
    1-\frac{d}2;\,1-v^2\Big),                      \label{6.1}
    \end{eqnarray}
where $F(a,b;c;u)$ is a hypergeometric function given by (\ref{A4}),
and its argument $u$ is defined in terms of the following two
functions of $\varepsilon=m/2M$,
    \begin{eqnarray}
    v_\pm\equiv\pm\sqrt{1+\varepsilon^2}
    -\varepsilon=-1/v_\mp.                         \label{6.2}
    \end{eqnarray}
This representation exactly recovers the nonanalytic part of the
$1/M$-expansion (\ref{3.20}), whereas the analytic part of the
latter follows from the known hypergeometric series
$F(a,b;c;u)=1+O(u)$ in powers of the argument
$u=1-v_\pm^2\sim\varepsilon$.

In the opposite limit of large $\varepsilon$ (corresponding to large
$m$ or small $M$) the brane action can be equivalently represented
in a form useful for the expansion in $v_+\sim 1/2\varepsilon\ll
1$. As shown in Appendix B by using the transformation formulae for
$F(a,b;c;u)$ it reads
    \begin{eqnarray}
    &&\frac12\;{\rm Tr}^{(d)}
    \ln \mbox{\boldmath$F$}^{\,\rm brane}=
    \left(\frac{M^2}{4\pi}\right)^{\!\frac{d}2}\int
    d^dx\,\left\{
    -\frac14\,\Gamma(-d/2)\,(1-v_+^2)^{d/2}\right.\nonumber\\
    &&\qquad\qquad\qquad\qquad\qquad\qquad
    +\frac{\pi}{2\,\Gamma(1+\frac{d}2)\,\sin(\pi d)}\,
    \frac{(1-v_+^2)^{d/2}}{v_+^d}              \nonumber\\
    &&\qquad\qquad\qquad\qquad\qquad\qquad
    +\frac{\Gamma\big(\!-\frac{1+d}2\big)}{4\,\sqrt\pi}\;
    v_+\left[\;(d+1)\,
    F\Big(1,\,\frac{1-d}2;\,
    \frac32;\,v_+^2\Big)\right.\nonumber\\
    &&\qquad\qquad\qquad\qquad\qquad\qquad\left.\left.-
    F\Big(1,\,\frac12;\,
    \frac{3+d}2;\,v_+^2\Big)\;\right]\right\}.     \label{6.3}
    \end{eqnarray}

Using this representation one can consider the limit of exactly
Neumann boundary conditions corresponding to $m\to\infty$ and a
finite value of the mass $M$ in the bulk. In this limit $v_+\to 0$
and the last two terms of (\ref{6.3}) vanish. Naively, the second
term $\sim\varepsilon^d$ is growing to infinity, but in the domain
of ultraviolet convergence $d<0$, so that with the appropriate order
of taking the limits (first in $m\to\infty$ and second in the
dimensionality) it vanishes, too. Thus, only the first term remains,
    \begin{eqnarray}
    &&\frac12\;{\rm Tr}^{(d)}
    \ln \mbox{\boldmath$F$}^{\,\rm brane}=-\frac14\,\Gamma(-d/2)\,
    \left(\frac{M^2}{4\pi}\right)^{\!\frac{d}2}\int
    d^dx,
    \end{eqnarray}
which is obviously one-half of the contribution of the
$d$-dimensional massive particle corresponding to the brane-to-brane
mode propagating with the square-root operator
$\mbox{\boldmath$F$}^{\,\rm brane}=\sqrt{M^2-\Box}$,
    \begin{eqnarray}
    &&\frac12\;{\rm Tr}^{(d)}
    \ln \mbox{\boldmath$F$}^{\,\rm brane}=\frac14\,{\rm Tr}^{(d)}
    \ln (M^2-\Box).
    \end{eqnarray}
When added to the Dirichlet effective action (\ref{2.5}) it alters
the sign of the brane ($d$-dimensional integral) term, which
corresponds to the transition from the Dirichlet value of the
$b_{1/2}^D$ surface coefficient (\ref{b1/2}) to the Neumann value,
    \begin{eqnarray}
    b_{1/2}^N=\frac{\sqrt\pi}2.              \label{b1/2N}
    \end{eqnarray}

Another interesting limit of $\varepsilon\to\infty$ corresponds to
the massless case of the DGP model with $M=0$. In this case the only
nonvanishing term is contained in the second line of (\ref{6.3}),
and it yields
    \begin{eqnarray}
    &&\frac12\;{\rm Tr}^{(d)}
    \ln \mbox{\boldmath$F$}^{\,\rm brane}=\frac12\,
    \left(\frac{m^2}{4\pi}\right)^{\!\frac{d}2}\,
    \frac{\pi}{\Gamma(1+\frac{d}2)\,\sin \pi d}\,
    \int d^dx,
    \end{eqnarray}
because $M/v_+\to 2\varepsilon M=m$. This result can be confirmed by
a direct calculation of the effective potential of the brane mode,
circumventing the operation of taking the limit $M\to 0$ in the
general answer (\ref{6.3}).

Indeed, the effective potential $V_{\rm eff}(m)$,
    \begin{eqnarray}
    \frac12\;{\rm Tr}^{(d)}
    \ln \mbox{\boldmath$F$}^{\,\rm brane}
    =\int d^d x\,V_{\rm eff}(m),
    \end{eqnarray}
for the brane operator in the toy DGP model,
$\mbox{\boldmath$F$}^{\,\rm brane}=(-\Box+m\sqrt{-\Box})/m$, can be
written down in the form of a momentum-space integral similar to
(\ref{2.6}),
    \begin{eqnarray}
    &&V_{\rm eff}(m)=\frac1{(4\pi)^{d/2}\,
    \Gamma(\frac{d}2)}\int\limits_0^\infty
    dp\,p^{d-1}\,\ln\,(p^2+m p)\nonumber\\
    &&\qquad\quad=
    \Big(\,\frac{m^2}{4\pi}\,
    \Big)^{\!d/2}\frac1{\Gamma(\frac{d}2)}\,
    \int\limits_0^\infty dt\,t^{d-1}\,
    \left[\,\ln\,(t+1)+\ln\,t+\ln\,m^2\,\right]\ .
    \end{eqnarray}
Within the dimensional regularization only the first logarithmic
term gives a nonvanishing contribution which we transform by using
the proper-time representation of the logarithm and changing the
order of integrations,
    \begin{eqnarray}
    &&V_{\rm eff}(m)=
    -\Big(\,\frac{m^2}{4\pi}\,
    \Big)^{\!d/2}\frac1{\Gamma(\frac{d}2)}\,
    \int\limits_0^\infty dt\,t^{d-1}\,
    \int\limits_0^\infty \frac{ds}s\,e^{-st-s}\nonumber\\
    &&\qquad\qquad\qquad\qquad\qquad\qquad\qquad\qquad
    =\frac12\,
    \Big(\,\frac{m^2}{4\pi}\,
    \Big)^{\!d/2}\frac1{\Gamma(\frac{d}2+1)}\,
    \frac\pi{\sin \pi d}.                               \label{6.10}
    \end{eqnarray}

After subtracting the ultraviolet divergence in the limit of the
physical dimensionality, $d\to N$, this gives the renormalized
effective potential of the usual Coleman-Weinberg structure,
    \begin{eqnarray}
    &&V_{\rm eff}(m)=\frac12\,\frac{(-1)^N}{\Gamma(\frac{N}2+1)}\,
    \Big(\,\frac{m^2}{4\pi}\,
    \Big)^{\!N/2}
    \ln\frac{m}{\mu}.                               \label{6.11}
    \end{eqnarray}
Here $\mu$ is the parameter reflecting the renormalization
ambiguity, and the role of the field (argument of the potential) is
played by the scale $m$ of the brane term in the classical action
(\ref{1.1}) -- the analogue of the DGP scale. This result confirms
the boundary effective action calculation of \cite{Pujolas}.

\section{Conclusions}
In conclusion, we have derived the one-loop effective action in a
simplified model of brane-induced gravity which gives rise to
special boundary conditions involving second order tangential
derivatives. The main peculiarity of this action is the presence of
logarithmic ultraviolet divergences for both even and odd
dimensionalities of the spacetime. This is different from analogous
one-loop calculations in spacetimes without boundaries leading to
divergences only for even spacetime dimensionalities. The action has
a nontrivial analytic structure in the mass parameter --- for
generic $M$ and $m$ it is given by two representations in terms of
hypergeometric functions (\ref{6.1}) and (\ref{6.3}) and simplifies
for the case of the massless field in the bulk, $M=0$, to the form
(\ref{6.10}). After ultraviolet renormalization this gives rise to
the familiar logarithmic Coleman-Weinberg effective potential $\sim
\varphi^d\ln(\varphi^2/\mu^2)$ with the field $\varphi$ played by
the parameter $m$ in the boundary conditions (\ref{1.3}) --- the
result used in \cite{Pujolas} for the stabilization of the DGP
cross-over scale in the Brans-Dicke modification of the DGP model
\cite{B-LopWands}.

We also derived the proper time expansion for the functional trace
of the heat kernel subject to these generalized Neumann boundary
conditions. This turns out to be nontrivial, because for even
dimensionalities of the boundary it involves together with the
well-known half-integer powers of the proper time $s$ also the
logarithmic terms $\sim\ln s$, cf. Eq.(\ref{HK3}), and for odd
dimensionalities contains powers of $s$ which are multiples of one
quarter, see Eq.(\ref{HK2}). Such peculiarities of the proper time
expansion are usually associated with the lack of strong ellipticity
of the boundary value problem \cite{ellipticity} when a naively
positive elliptic operator acquires due to the presence of the
boundary an infinite set of negative modes. These modes make the
heat kernel operator unbounded and violate usual assumptions
underlying its proper time expansion. But, as we show in Appendix C,
strong ellipticity of our problem gets violated only for negative
$m$ in (\ref{1.3}) --- when the relevant classical action
(\ref{1.1}) and the brane-to-brane operator
$\mbox{\boldmath$F$}^{\,\rm brane}$, Eq.~(\ref{9}), are both not
positive definite. However, the heat kernel expansions become exotic
also for $m>0$ with no violation of strong ellipticity, which
implies deeper reasons of these peculiarities.

To summarize, we conclude that the technique for quantum effects in
brane models is more complicated than in systems without boundaries.
Moreover, it does not reduce to a simple bookkeeping of surface
terms in the heat kernel expansion of
\cite{McKean-Singer,Osborn-McAvity,Vassilevich}, and so on, because
of the complicated square-root structure of the brane propagator
(\ref{9}) mediating the effect of the generalized Neumann boundary
conditions (\ref{1.3}). The proper time method that was
fundamentally efficient in models without boundaries
\cite{DeWitt,PhysRep} in our calculations above became a derivative
of an alternative calculation. Namely, the surface terms in the heat
kernel expansion were recovered from the $1/M$-expansion of the
action obtained by a different method of a Fourier decomposition.

Nevertheless, the proper time method still does not loose its power
and can be used in realistic brane models including gravity. In
these models the effective action should be expanded in powers of
the bulk spacetime curvature and the extrinsic curvature of the
brane, starting with the approximation considered above. The
momentum space decomposition is not very efficient for sake of such
an expansion --- the difficulty usually circumvented in background
field formalism by the use of the Schwinger-DeWitt proper time
method \cite{DeWitt,DeWitt1,PhysRep}. Here we present without
derivation (that will be given in forthcoming publications) such a
representation for the Green's function of the DGP brane-to-brane
operator and its functional determinant. They have a form of the
weighted proper time integrals
    \begin{eqnarray}
    &&\frac1{-\Box+m\sqrt{-\Box}}=
    \int\limits_0^\infty ds\,e^{s\,\Box}\,w(s),  \label{7.1}\\
    &&{\rm Tr}\,\ln\Big(\!-\Box+m\sqrt{-\Box}\,\Big)=
    -{\rm Tr}\,
    \int\limits_0^\infty\frac{ds}s\,
    e^{s\,\Box}\,\frac{1+w(s)}2,       \label{7.2}
    \end{eqnarray}
with the weight function $w(s)$ given in terms of the error function
$\Phi(x)=\frac2{\sqrt\pi}\int_0^x dy\,\exp(-y^2)$ and having the
following ultraviolet and infrared asymptotics
    \begin{eqnarray}
    w(s)=e^{sm^2}
    \Big(1-\Phi(\,m\sqrt{s}\,)\Big)
    \to\left\{\begin{array}{ll}\,1,\,\,\,\,\,&m\sqrt{s}\to 0,\\
    \,1/m\sqrt{s\pi},\,\,\,\,\,&\,m\sqrt{s}
    \to \infty.\end{array}\right.
    \end{eqnarray}
The advantage of this representation\footnote{Note, in passing, that
the interpretation of this weight and its asymptotics is very
transparent. In the ultraviolet domain of small proper time
$m\sqrt{s}\ll 1$ (or big $\sqrt{-\Box}\gg m$) it approximates the
brane operator by $-\Box$, whereas in the infrared domain
$m\sqrt{s}\gg 1$ (or $\sqrt{-\Box}\ll m$) it corresponds to its
low-energy behavior $m\sqrt{-\Box}$. All the results above could be
obtained with the aid of this representation generalized to the case
of a nonzero $M$. We did not use it, however, because this
generalization has a complicated weight function.} is that it
applies also to the case of the curved-space d'Alember\-tian $\Box$,
so that the generally covariant expansion of (\ref{7.1})-(\ref{7.2})
in curvatures can be directly obtained by using a well-known
Schwinger-DeWitt expansion for $e^{s\Box}$. Thus the lowest order
approximation for the exact brane-to-brane operator (\ref{5}) in
models with a curved bulk and curved branes can be considered by
means of the manifestly covariant technique which can be
systematically extended to higher orders. Combined with the method
of fixing the background covariant gauge for diffeomorphism
invariance in brane models, developed in \cite{qeastbg}, this will
ultimately give the universal background field method of the
Schwinger-DeWitt type in gravitational brane systems.

\appendix
\renewcommand{\thesection}{\Alph{section}.}
\renewcommand{\theequation}{\Alph{section}.\arabic{equation}}

\section{Asymptotic expansion for integrals with a weak singularity}
The analytic and nonanalytic parts of the asymptotic expansion for the
integral (\ref{3.4}) (a so-called integral with a weak
singularity) can be found by the method of \cite{Fedoryuk}. For an
integral of a slightly more general form this expansion reads
    \begin{eqnarray}
    &&\int\limits_0^\infty dt\,t^{\beta-1}\,
    (t+\varepsilon)^\alpha\,
    \varphi(t)=\varepsilon^{\alpha+\beta}\,
    \sum\limits_{n=0}^\infty
    \frac{\Gamma(n+\beta)\,
    \Gamma(-\alpha-\beta-n)}{\Gamma(-\alpha)}\,
    \frac{\varphi^{(n)}(0)}{n!}\,\varepsilon^n\nonumber\\
    &&\qquad\qquad\qquad\qquad\qquad\qquad\qquad\qquad
    +\sum\limits_{n=0}^\infty
    a_n\,\varepsilon^n,\,\,\,\,\varepsilon\to 0.         \label{3.7}
    \end{eqnarray}
Here $\varphi(t)$ is a function which is analytic at $t=0$ and has a
Taylor series with coefficients
$\varphi^{(n)}(0)\equiv d^n\varphi/dt^n(0)$. The parameter $\beta$ is
positive in order to guarantee the convergence of the integral at $t=0$. The
first sum gives a nonanalytic part of the expansion determined
entirely by the derivatives of the function $\varphi$, $\varphi^{(n)}(0)$,
whereas the second sum determines the analytic part with
coefficients involving a (nonlocal) dependence of the function
$\varphi(t)$ at all $t$. These coefficients are given by the
following expression:
    \begin{eqnarray}
    &&a_n=\frac{\alpha(\alpha-1)...(\alpha-n+1)}{n!}
    \left\{\;\int\limits_0^\infty dt\,
    \varphi_S(t)\,t^{\alpha+\beta-n-1}
    \right.\nonumber\\
    &&\qquad\qquad\qquad\qquad\qquad\qquad\qquad
    \left.+\sum_{m=0}^{[\,n-\alpha-\beta\,]}
    \frac{T^{\alpha+\beta+m-n}}{\alpha+\beta+m-n}
    \frac{\varphi^{(m)}(0)}{m!}\,
    \right\},                                 \label{3.8}
    \end{eqnarray}
where $\varphi_S(t)$ is a piecewise smooth function obtained from
$\varphi(t)$ by subtracting its first few terms of the Taylor
expansion at $t=0$ on a finite segment of the $t$-axes, $0\leq t<
T$,
    \begin{eqnarray}
    &&\varphi_S(t)=\varphi(t)-\!\!\sum_{m=0}^{[\,n-\alpha-\beta\,]}
    \frac{\varphi^{(m)}(0)}{m!}\,t^m,\,\,\,t<T,\nonumber\\
    &&\varphi_S(t)=\varphi(t),\,\,\,\,\,\,\,\,\,t\geq T.   \label{3.9a}
    \end{eqnarray}
Here the number of subtracted terms is given by
$[\,n-\alpha-\beta\,]$ --- the integer part of $n-\alpha-\beta$, and
$T$ is arbitrary positive. The value of the latter is immaterial,
because it is easy to check that $\partial a_n/\partial T=0$. These
subtractions are necessary to guarantee the convergence of the
integrals in (\ref{3.8}) at $t=0$. For the first few $a_n$ these
subtractions are absent,
    \begin{eqnarray}
    a_n=\frac{\alpha(\alpha-1)...(\alpha-n+1)}{n!}
    \int\limits_0^\infty dt\,
    \varphi(t)\,t^{\alpha+\beta-n-1},\,\,\,
    n<\alpha+\beta,                    \label{3.9}
    \end{eqnarray}
while for $n>\alpha+\beta$ their effect can be explicitly
circumvented by multiple integrations by parts in (\ref{3.8}). After
$(N_n+1)$ integrations by parts, all non-integral terms of
(\ref{3.8}) cancel out and the expansion coefficients take the
following alternative form,
    \begin{eqnarray}
    &&a_n=\frac{\alpha(\alpha-1)...(\alpha-n+1)}{n!}\,
    \frac{\Gamma(n-\alpha-\beta-N_n+1)}
    {\Gamma(n-\alpha-\beta+1)}\nonumber\\
    &&\qquad\qquad\qquad
    \times\,\frac1{n-\alpha-\beta-N_n}\,
    \int\limits_0^\infty dt\;
    \varphi^{(N_n+1)}(t)\,
    t^{\alpha+\beta-n+N_n},\,\,\,
    n>\alpha+\beta,                              \label{3.10}
    \end{eqnarray}
where
    \begin{eqnarray}
    N_n=[\,n-\alpha-\beta\,].             \label{3.11}
    \end{eqnarray}
Finally, when $\alpha+\beta$ is a positive integer $N$ the two sums
of (\ref{3.7}) formally become analytic, but their coefficients
develop pole singularities in $\alpha+\beta-N\to 0$. These
singularities come from these two sums with opposite signs and
cancel. The finite remnant of this cancellation is a term logarithmic in
$\varepsilon$.

If we apply now this asymptotic expansion to the case of our
integral (\ref{3.3})--(\ref{3.4}) with
    \begin{eqnarray}
    \varphi(t)=(\sqrt{1+t^2}+t)^{d/2},\,\,\,\,
    \alpha=-1,\,\,\,\,\beta=\frac{d}2\ ,   \label{3.12}
    \end{eqnarray}
then in view of
    \begin{eqnarray}
    \frac{\Gamma(n+\beta)\,
    \Gamma(-\alpha-\beta-n)}{\Gamma(-\alpha)}
    =(-1)^n\frac\pi{\sin(\pi d/2)} \ ,                 \label{3.13}
    \end{eqnarray}
the nonanalytic part is explicitly expressed in terms of the
function $\varphi(-\varepsilon)$ with the flipped sign of its
argument,
    \begin{eqnarray}
    &&\frac{\pi\,\varepsilon^{d/2-1}}{\sin(\pi d/2)}\,
    \sum\limits_{n=0}^\infty
    \frac{\varphi^{(n)}(0)}{n!}\,(-\varepsilon)^n=
    \frac{\pi\,\varepsilon^{d/2-1}\,
    \varphi(-\varepsilon)}{\sin(\pi d/2)}\ .        \label{3.14}
    \end{eqnarray}

The coefficients of the analytic part can be explicitly calculated
by taking the integrals (\ref{3.9}) in the domain $n<d/2-1$ and
extended beyond this domain by analytic continuation (which is
equivalent to using (\ref{3.10})). The result is
    \begin{eqnarray}
    a_n=2^{n-1-d/2}(-1)^{n+1}\,d\,
    \frac{\Gamma(\frac{n-d+1}2)\,
    \Gamma(\frac{d}2-1-n)}{\Gamma(-\frac{n-1}2)}.     \label{3.16}
    \end{eqnarray}
Therefore for odd $n=2j+1$ they vanish due to the unregulated (by
the dimensionality $d$) infinity in the denominator, whereas for
even $n=2j$ they are given by (\ref{3.18}), and the final form of
the expansion for $I$ is given by (\ref{3.14one}).

\section{Hypergeometric function representation}
The integral (\ref{2.6}) can be rewritten in terms of the
integration variable $x=\sqrt{p^2+M^2}/M$ as
    \begin{eqnarray}
    &&\int\limits_0^\infty dp\,p^{d-1}
    \,\ln \left(p^2+m\sqrt{M^2+p^2}\,\right)\nonumber\\
    &&\qquad\qquad\qquad\qquad\quad
    =M^{d}\,\int\limits_{1}^{\infty}
    d x\; x \,(x^2-1)^{d/2\!-\!1} \ln{\Big(x^2
    +2\varepsilon x-1\Big)}.           \label{999}
    \end{eqnarray}
By factorizing the argument of the logarithm and integrating the
result by parts we convert it to the sum of two terms,
    \begin{eqnarray}
    &&\int\limits_{1}^{\infty}
    d x\; x \,(x^2-1)^{d/2\!-\!1} \ln{\Big(x^2
    +2\varepsilon x-1\Big)}\nonumber\\
    &&\qquad\qquad\qquad\qquad
    =\sum\limits_{v\,=\,v_\pm}\,
    \int\limits_{1}^{\infty} d x\;
    x \,(x^2-1)^{d/2\!-\!1}
    \ln{(x-v)}\nonumber\\
    &&\qquad\qquad\qquad\qquad=
    -\frac1d\,\sum\limits_{v\,=\,v_\pm}\,
    \int\limits_{1}^{\infty}
    d x\;(x^2-1)^{d/2}\,(x-v)^{-1},     \label{1000}
    \end{eqnarray}
where $v_\pm$ are the roots (\ref{6.2}) of the quadratic polynomial
$x^2+2\varepsilon x-1=(x-v_+)(x-v_-)$.

With the change of the integration variables $x=1/\sqrt{t}$ we have
    \begin{eqnarray}
    &&\int\limits_{1}^{\infty} d x\;
    (x^2-1)^{d/2}\,(x-v)^{-1}=\frac12\,\int\limits_{0}^{1} dt\;
    t^{-1-d/2}(1-t)^{d/2}\,(1-t v^2)^{-1}\nonumber\\
    &&\qquad\qquad\qquad\qquad\quad+
    \frac{v}2\,\int\limits_{0}^{1} dt\;t^{-1/2-d/2}\,
    (1-t)^{d/2}\,(1-t v^2)^{-1},
    \end{eqnarray}
so that finally in terms of the hypergeometric function
    \begin{eqnarray}
     F(a,b;c;u)=\frac{\Gamma(c)}{\Gamma(b)\Gamma(c-b)}
     \int_0^1 dt\,t^{b-1}\,(1-t)^{c-b-1}\,
     (1-t\,u)^{-a}                       \label{A4}
    \end{eqnarray}
the basic integral reads as
    \begin{eqnarray}
    &&\int\limits_{1}^{\infty} d x\;
    (x^2-1)^{d/2}\,(x-v)^{-1}=\frac{\Gamma(-{d\over2})\,
    \Gamma(1+{d\over2})}{2\Gamma(1)}\,
    F\Big(1,-\frac{d}2;\,1;\,v^2\Big)\nonumber\\
    &&\qquad\qquad\qquad\qquad\qquad\qquad
    +\frac{v}2\,\frac{\Gamma(1+{d\over2})\,
    \Gamma(\frac{1-d}2)}{\Gamma({3\over2})}\,
    F\Big(1,\,\frac{1-d}2;\,\frac32;\,v^2\Big)\ .   \label{A11}
    \end{eqnarray}

We will need its $\varepsilon$-expansion. At $\varepsilon\to 0$ the
parameter $v=v_\pm\to \pm 1$, so we have to transform the
hypergeometric functions to the series in $1-v^2$. Because of the
known relation $F(a,b;a;u)=F(b,a;a;u)=(1-u)^{-b}$ we have
    \begin{eqnarray}
     F\Big(1,-\frac{d}2;\,1;\,v^2\Big)=(1-v^2)^{d/2}  \label{A12}
    \end{eqnarray}
and in view of Eq. 9.131.2 of \cite{GR},
    \begin{eqnarray}
     &&F\Big(1,\,\frac{1-d}2;\,\frac32;\,v^2\Big)=
     \frac{\Gamma({3\over2})\,\Gamma({d\over2})}
     {\Gamma({1\over2})\,\Gamma(1+{d\over2})}\,
    F\Big(1,\frac{1-d}2;\,1-{d\over2};\,1-v^2\Big)\nonumber\\
    &&\qquad\qquad\qquad\qquad\quad
    +(1-v^2)^{d/2}\,\frac{\Gamma({3\over2})\,
    \Gamma(-\frac{d}2)}{\Gamma(1)\,\Gamma(\frac{1-d}2)}\,
    F\Big(\,{1\over2},\,1+\frac{d}2;\,1+\frac{d}2;\,1-v^2\Big),
    \end{eqnarray}
where again
$F\Big(\,{1\over2},\,1+\frac{d}2;\,1+\frac{d}2;\,1-v^2\Big)
=[1-(1-v^2)]^{-1/2}=|v|^{-1}$. Therefore
    \begin{eqnarray}
     &&F\Big(1,\,\frac{1-d}2;\,\frac32;\,v^2\Big)=
     -\frac{\sqrt\pi}d\,\frac{\Gamma(1-\frac{d}2)}{\Gamma(\frac{1-d}2)}\,
     \frac{(1-v^2)^{d/2}}{|v|}\nonumber\\
     &&\qquad\qquad\qquad\qquad\qquad\qquad\qquad\qquad+
    \frac1d\,
    F\Big(1,\frac{1-d}2;\,1-{d\over2};\,1-v^2\Big)  \label{A14}
    \end{eqnarray}

Substituting (\ref{A12}) and (\ref{A14}) into (\ref{A11}) we have
    \begin{eqnarray}
    &&\int\limits_{1}^{\infty} d x\;
    (x^2-1)^{d/2}\,(x-v)^{-1}=
    -\frac{\pi}{\sin(\pi d/2)}\,\theta(v)\,(1-v^2)^{d/2}   \nonumber\\
    &&\qquad\qquad\qquad\qquad\quad
    +\frac1{2\sqrt\pi}\,\Gamma\Big(\,{d\over2}\,\Big)\,
    \Gamma\Big(\,\frac{1-d}2\,\Big)\,
    v\,F\Big(1,\,\frac{1-d}2;\,1-\frac{d}2;\,1-v^2\Big),   \label{100000}
    \end{eqnarray}
where a step function $\theta(v)$ arose as the result of summation
of two terms,
    \begin{eqnarray}
    \theta(v)=\frac12\,\Big(\,1+\frac{v}{|v|}\,\Big).
    \end{eqnarray}
The first term in (\ref{100000}) exists only for positive $v$ and is
nonanalytic at $v\to 1$, whereas the second term is analytic.
Obviously for negative $v$ this integral is an analytic function
because the argument of the logarithm nowhere tends to zero in the
integration domain. This explains the absence of the first term for
$v<0$.

Substituting (\ref{100000}) to (\ref{999})-(\ref{1000}) we finally
get
    \begin{eqnarray}
    &&\int\limits_0^\infty dp\,p^{d-1}
    \,\ln \left(p^2+m\sqrt{M^2+p^2}\,\right)=
    \frac{M^d}d\,\frac{\pi}{\sin(\pi d/2)}\,(1-x_+^2)^{d/2}   \nonumber\\
    &&\qquad\qquad\qquad\qquad
    -\frac{M^d}d\,\frac{\Gamma\big(\,{d\over2}\,\big)\,
    \Gamma\big(\,\frac{1-d}2\,\big)}{2\,\sqrt\pi}
    \,\sum\limits_{v\,=\,v_\pm}\,
    v\,F\Big(1,\,\frac{1-d}2;\,1-\frac{d}2;\,1-v^2\Big).     \label{A16}
    \end{eqnarray}
Bearing in mind that $(1-v_+^2)^{d/2}=(2\varepsilon
v_+)^{d/2}=(2\varepsilon)^{d/2}\varphi(-\varepsilon)$ we finally
have the representation useful for small $\varepsilon$,
    \begin{eqnarray}
    &&\int\limits_0^\infty dp\,p^{d-1}
    \,\ln \left(p^2+m\sqrt{M^2+p^2}\,\right)=
    \frac{M^d}d\,\frac{\pi}{\sin(\pi d/2)}\,
    (2\varepsilon)^{d/2}\varphi(-\varepsilon)\nonumber\\
    &&\qquad\qquad\qquad\qquad
    -\frac{M^d}d\,\frac{\Gamma\big(\,{d\over2}\,\big)\,
    \Gamma\big(\,\frac{1-d}2\,\big)}{2\,\sqrt\pi}
    \,\sum\limits_{v\,=\,v_\pm}\,
    v\,F\Big(1,\,\frac{1-d}2;\,1-\frac{d}2;\,1-v^2\Big),     \label{17}
    \end{eqnarray}
which gives rise to the representation (\ref{6.1}).

To consider the limit of $\varepsilon\to\infty$ we need another
representation, because in this limit $v_+\to 1/2\varepsilon$ and
$v_-\sim -2\varepsilon\to -\infty$, so that the contribution of
$v=v_-$ in (\ref{1000}) should be expandable in $1/v$. This can be
achieved by the transformation formula 9.132.2 of \cite{GR} which in
our case reads as
    \begin{eqnarray}
     &&F\Big(1,\,\frac{1-d}2;\,\frac32;\,v^2\Big)=
     \frac1{1+d}\,\frac1{v^2}\,
    F\Big(1,\,\frac12;\,{{3+d}\over2};\,{1\over{v^2}}\Big)\nonumber\\
    &&\qquad\qquad\qquad\qquad\qquad\qquad\qquad\qquad
    +\frac{\sqrt\pi}2\,\frac{\Gamma({{1+d}\over2})}{\Gamma(1+\frac{d}2)}\,
    \frac{(1-v^2)^{d/2}}{(-v^2)^{1/2}}\ .
    \end{eqnarray}
Using this in the representation (\ref{11}) of the $v=v_-$ term of
(\ref{1000}) and taking into account that $1/v_-=-v_+$ we finally
arrive at the equation underlying the representation (\ref{6.3}):
    \begin{eqnarray}
    &&\int_0^\infty dp\,p^{d-1}
    \,\ln \left(p^2+m\sqrt{M^2+p^2}\,\right)\nonumber\\
    &&\qquad\qquad\qquad\quad
    =\frac{\pi\,M^d}{2 d\,\sin(\pi d/2)}\,(1-v_+^2)^{d/2}
    +
    \frac{\pi\,M^d\,}{d\,\sin(\pi d)}\,\frac{(1-v_+^2)^{d/2}}{v_+^d}\nonumber\\
    &&\qquad\qquad\qquad\quad
    +M^d\,\frac{\Gamma\big(\,{d\over2}\,\big)\,
    \Gamma\big(-\frac{1+d}2\big)}{4\,\sqrt\pi}\;v_+\left[\;(d+1)\,
    F\Big(1,\,\frac{1-d}2;\,\frac32;\,v_+^2\Big)\right.\nonumber\\
    &&\qquad\qquad\qquad\qquad\qquad\qquad\qquad\qquad
    \qquad\qquad\qquad\left.-
    F\Big(1,\,\frac12;\,\frac{3+d}2;\,v_+^2\Big)\;\right].
    \end{eqnarray}

\section{Strong ellipticity problem}
The strong ellipticity problem for the operator (\ref{6}) with
generalized Neumann boundary condition (\ref{1.3}) consists in the
existence of an infinite set of normalizable eigenmodes with a
spectrum which is
unbounded from below \cite{ellipticity}. This implies that
the relevant heat kernel is an unbounded operator which cannot be
rendered bounded by the elimination of a finite number of states
from its functional space, and therefore it has an unusual proper
time asymptotics different from (\ref{10}). This situation occurs
when the parameter $m$ in the operator (\ref{8}) of the boundary
condition is negative\footnote{We are grateful to D.Vassilevich for
pointing out this example of strong ellipticity violation.}. Here
we show that these negative modes correspond to the negative modes
of the brane-to-brane operator (\ref{9}), localized near the
brane/boundary and responsible for brane-to-brane propagation.

Indeed, in this case there is a set of eigenmodes localized near the
boundary $y=0$ of the form
    \begin{eqnarray}
    \Phi_p(x,y)=e^{ipx-\lambda_p y}
    \end{eqnarray}
in which $\lambda_p$ is given on account of the boundary condition
by the expression
    \begin{eqnarray}
    \lambda_p=-\frac{p^2}m>0,\,\,\,\,\,m<0.
    \end{eqnarray}
Negative $m$ guarantees the normalizability of these eigenmodes
which, therefore, cannot be excluded from the functional space of
the operator. Their eigenvalues $\Lambda_p$,
    \begin{eqnarray}
    &&(M^2-\Box_{(d+1)})\,\Phi_p(X)=
    \Lambda_p\,\Phi_p(X),                       \\
    &&\Lambda_p=
    \Big(M^2+p^2-\frac{(\,p^2\,)^2}{m^2}\,\Big),
    \end{eqnarray}
are negative for sufficiently high Fourier momenta $p$,
    \begin{eqnarray}
    \Lambda_p<0,\,\,\,\,\,p^2>\frac{m^2}2\,
    \Big(1+\sqrt{1+4M^2/m^2}\,\Big),
    \end{eqnarray}
and tend to $-\infty$ for $p^2\to\infty$. But the momentum space
domain where they are negative exactly coincides with the domain in
which the brane operator is negative definite for $m<0$,
    \begin{eqnarray}
    \mbox{\boldmath$F$}^{\,\rm brane}\,\varphi_p(x)=
    \left(\frac{p^2}{m}+\sqrt{M^2+p^2}
    \right)\varphi_p(x),\,\,\,\,\varphi_p(x)=e^{ipx}.
    \end{eqnarray}

Thus, the lack of strong ellipticity of the generalized Neumann
boundary value problem is in fact the lack of positivity of the
action (\ref{1.1}) with $m<0$, from which this problem originates by
the variational procedure.

\section*{Acknowledgements}
A.B. thanks D.Vassilevich for helpful thought provoking discussions.
A.B. and A.Yu.K. are grateful for the hospitality of the Institute
for Theoretical Physics at the University of Cologne where a major
part of this work has been done under the grant 436 RUS 17/8/06 of
the German Science Foundation (DFG). A.B. was partially supported by
the RFBR grant No 05-01-00996 and the LSS grant No 4401.2006.2. A.K.
was partially supported by the RFBR grant 05-02-17450 and the LSS
grant 1757.2006.2. D.N. was supported by the RFBR grant No
05-02-17661 and the LSS grant No 4401.2006.2 and also thanks the
Center of Science and Education of the Lebedev Institute and the
target funding program of the Presidium of Russian Academy of
Sciences for support. At the completing stage of this work A.B. was
also supported by the SFB 375 grant at the Physics Department of the
Ludwig-Maximilians University in Munich.

\end{document}